\documentclass[aps,prb,twocolumn]{revtex4}
\usepackage{graphicx}
\usepackage{hyperref}
\usepackage{amsmath}

\begin{document}
\title{Magnetic field-enhanced spin filtering in rare-earth mononitride tunnel junctions}
\author{P. K. Muduli$^1$, X. L. Wang$^2$, J. H. Zhao$^2$, Mark G. Blamire$^1$}

\affiliation{$^1$Department of Materials Science and
Metallurgy,University of Cambridge, 27 Charles Babbage Road,
Cambridge CB3 0FS,United Kingdom}
\email{pkm27@cam.ac.uk}\affiliation{$^2$State Key Laboratory of
Superlattices and Microstructures, Institute of Semiconductors,
Chinese Academy of Sciences, Beijing 100083, China}

\begin{abstract}
Spin filter tunnel junctions are based on selective tunneling of
up and down spin electrons controlled through exchange splitting
of the band structure of a ferromagnetic insulator. Therefore,
spin filter efficiency can be tuned by adjusting exchange strength
of the tunnel barrier. We have observed that magnetic field and
bias voltage (current) can be used to  regulate exchange strength
and consequently spin-filter efficiency in tunnel junctions with
ferromagnetic DyN and GdN tunnel barrier. In tunnel junctions with
DyN barrier we obtained $\sim$37$\%$ spin polarization of
tunneling electrons at 11 K due to a small exchange splitting  ($
E_{ex}$) $\approx$5.6 meV of the barrier height ($\Phi _0$)
$\approx$60 meV. Huge spin-filter efficiency $\sim$97$\%$ was
found for tunnel junctions with GdN barrier due to larger $E_{ex}$
$\approx$47 meV. In the presence of an applied magnetic field,
barrier height can further split  due to magnetic field dependent
exchange splitting $ E_{ex}(H)$. The spin filter efficiency in DyN
tunnel junctions can be increased up to $\sim$87$\%$ with magnetic
field. Electric and magnetic field tuned spin-filter efficiency of
these tunnel junctions gives opportunity for practical application
of these devices with additional functionality.
\end{abstract}

\maketitle

\section{Introduction}

The creation, injection and transport of spin-polarized current
are the three basic  steps  in all spintronics devices. Instead of
relying on conventional ferromagnets for creation, fully spin
polarized currents can also be generated using spin-filter
tunneling of electrons through a ferromagnetic tunnel barrier. The
spin-up and spin-down electrons of a  non-magnetic electrode
passing through a spin-filter tunnel barrier can be filtered due
to the difference in tunnel barrier heights for the two spin
channels\cite{miao,moderarev}. The difference in barrier height
appears due to exchange splitting of the band structure, which
leads to the conduction band minima and valance band maxima at
different energies for majority and minority spin electrons. Spin
filter tunnel barrier can also solve impedance mismatch problem
with semiconducting counter electrodes facilitating in spin
injection from even a nonmagnetic metal into a
semiconductor\cite{schmidt}. Many ferromagnetic insulators have
been tested for their spin filtering property: the Eu
chalcogenides including EuS \cite{moodera-hao,hao}, EuSe
\cite{moodera}, EuO \cite{santos}, etc. have shown high spin
filtering efficiency; oxides like
CoFe$_2$O$_4$\cite{chapline,ramos}, NiFe$_2$O$_4$\cite{luders},
BiMnO$_3$\cite{gajek},
NiMn$_2$O$_4$\cite{nelson},CoCr$_2$O$_4$\cite{chopdekar} and
Sm$_{0.75}$Sr$_{0.25}$MnO$_3$\cite{prasad}, etc. have also
received a lot of attention due to their higher Curie temperature,
but to date, the spin filter efficiencies have been low.

Many of the rare earth mono nitrides (REN) are magnetic
semiconductors in which magnetic and electronic properties are
strongly coupled\cite{natali,cgduan}. Magnetism in REN is very
complex and not yet fully understood; some of these materials
display a peculiar type of forced ferromagnetism called
metamagnetism\cite{mayer}; i.e., they are antiferromagnetic at low
magnetic field and forced to be ferromagnetic by applying high
magnetic field. Usually RENs undergo a second order magnetic
transition from paramagnetic semiconducting behavior to a
ferromagnetic metallic-like state below the Curie temperature
($T_{Curie}$). As $4f$ moments in RE atoms are highly localized,
the exchange interaction in these compounds is determined by
indirect exchange interaction between these localized moments.
Most widely studied feromagnetic REN is GdN which shows the
highest $T_{Curie}$ $\sim$70 K. Recently, GdN has shown spin
filtering effect in NbN-GdN-NbN tunnel
junctions\cite{senapati,blamire,pal}. This opens up rare earth
mono nitrides as another class of spin-filter tunnel barriers. In
this paper, we have investigated the spin filtering property of
poorly explored DyN and GdN tunnel barriers. DyN is a
ferromagnetic semiconductor with $T_{Curie}$ $\sim$35
K\cite{muduli}. DyN has two more $4f$ electrons than the
half-filled GdN and it has the highest saturation magnetization
$\sim$10$\mu_B$/Dy among the REN series. Theoretical band
structure calculations on DyN show a small indirect $\Gamma$-$X$
gap $\sim$0.34 eV and a minimum direct gap of $\sim$1.17 eV at
$X$\cite{larson}. Experimentally DyN has been shown to be a
semiconductor with an optical gap of $\sim$1.2 eV\cite{azeem}.
Recently, we have made an extensive study of the tunneling
property of NbN-DyN-NbN tunnel junctions with different thickness
of DyN\cite{muduli}. We found a crossover from diffusive to
tunneling  transport as the DyN thickness is made smaller than
$\sim$4 nm. In this paper, we  show that the tunneling properties
of DyN and GdN junctions are strongly affected by magnetic field.
We demonstrate that splitting of the tunnel barrier height for
spin-up and down electron can be further increased with magnetic
field. We have explained our data with simple tunneling model of
Simmons considering different conductance channel for up-spin and
down-spin electrons.

\section{Experiment}

NbN-DyN-NbN and NbN-GdN-NbN trilayer films were deposited on 5
$\times$ 5 mm$^2$ Si/SiO$_2$(250 nm)/MgO(10 nm) substrates by
reactive DC sputtering. The deposition was done at room
temperature with a pressure of $\sim$1.5 Pa  in an UHV chamber
equipped with multiple layer deposition. Top and bottom 50 nm
thick superconducting NbN was deposited with 28$\%$ Ar-N$_2$ gas
mixture at 100 W sputtering power. Whereas 8$\%$ Ar-N$_2$ gas
mixture with lower $\sim$16 and $\sim$20 W sputtering power was
used for DyN and GdN, respectively. The details of deposition
condition is described in reference\cite{muduli}. Trilayers with
2.6 nm GdN and 3 nm DyN barriers were used for tunnel junction
fabrication. Eight tunnel junctions of dimension 7 $\times$ 7
$\mu$m$^2$ were fabricated on each substrate by optical
lithography in conjunction with Ar-ion milling and CF$_4$ plasma
etching. Electrical characterization of the tunnel junctions were
done in a four-probe configuration with a closed-cycle He
refrigerator from Cryogenic Limited. Although all the junctions
behaved in a similar way, for consistency we have shown
measurements done on one junction of each type only in this paper.
Magnetization measurements were performed in a Quantum Design
SQUID magnetometer.

\section{Results}
\begin{figure}[!h]
\begin{center}
\abovecaptionskip -10cm
\includegraphics [width=8 cm]{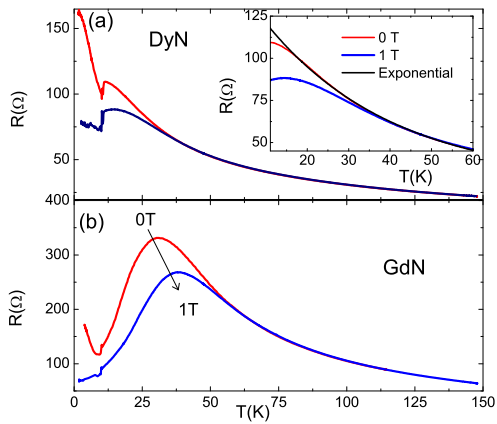}
\end{center}
\caption{\label{fig1} (Color online) Temperature dependence of
resistance of (a) NbN-DyN(3 nm)-NbN  and (b) NbN-GdN(2.6 nm)-NbN
tunnel junction with zero (red) and 1 T (blue) magnetic field. The
measurement was done with a bias current $I$ = 1 $\mu$A.}
\end{figure}
Figure 1 shows the junction resistance as a function of
temperature $R(T)$ with  magnetic field $\mu_0H$ = 0 and 1 T.
Semiconducting behavior can be seen in $R(T)$ above 11 and 50 K
for the DyN and GdN junction, respectively. Evidence for spin
filtering is obtained from $R(T)$ below $T_{Curie}$ when exchange
splitting of the barrier reduces the barrier height for one spin
sign and hence a decrease in the resistance of the junction.
Effect of spin filtering can be seen clearly as drop in $R(T)$
below $\sim$30 K in the case of the GdN junction as shown in Fig.
1(b). In the case of the DyN junction the deviation of the
experimental $R(T)$ data below $\sim$15 K from the extrapolated
exponential fit above $\sim$20 K provides proof that spin
filtering is present in these junctions. The spin filtering
efficiency can be determined from $R(T)$ using the method
described in the reference\cite{blamire}. In the low bias limit
spin-filter efficiency can be written as, $P \approx \sqrt {1 -
R_r^2 }$, where $R_r$ is junction resistance relative to the
non-magnetic state. For the DyN junction spin filtering efficiency
$P$ $\sim$37$\%$ was found at 11 K in zero magnetic field. This is
quite small compared to $P$ $\sim$97$\%$ found at 11 K for the GdN
tunnel junction. When an in-plane magnetic field is applied we
found that the spin filter efficiency increased from $P$ $\sim$67
$\%$ at 1 T to $P$ $\sim$87 $\%$ at 5 T for the DyN tunnel
junction. A small increase of $P$ $\sim$97 $\%$ at zero field to
$P$ $\sim$98 $\%$ at 1 T was found for the GdN junction at the
same temperature. The change in slope in the $R(T)$ ($dR(T)/dT >
0$ to $dR(T)/dT < 0$) can be used as an indicator of $T_{Curie}$.
Fig. 1(b) shows that for the GdN junctions, the $T_{Curie}$ is
significantly enhanced from  30 to 38 K when 1 T magnetic field
was applied. Similar enhancement of $T_{Curie}$ can also be seen
for the DyN junctions in Fig. 1(a).

\begin{figure}[!h*]
\begin{center}
\abovecaptionskip -10cm
\includegraphics [width=9 cm]{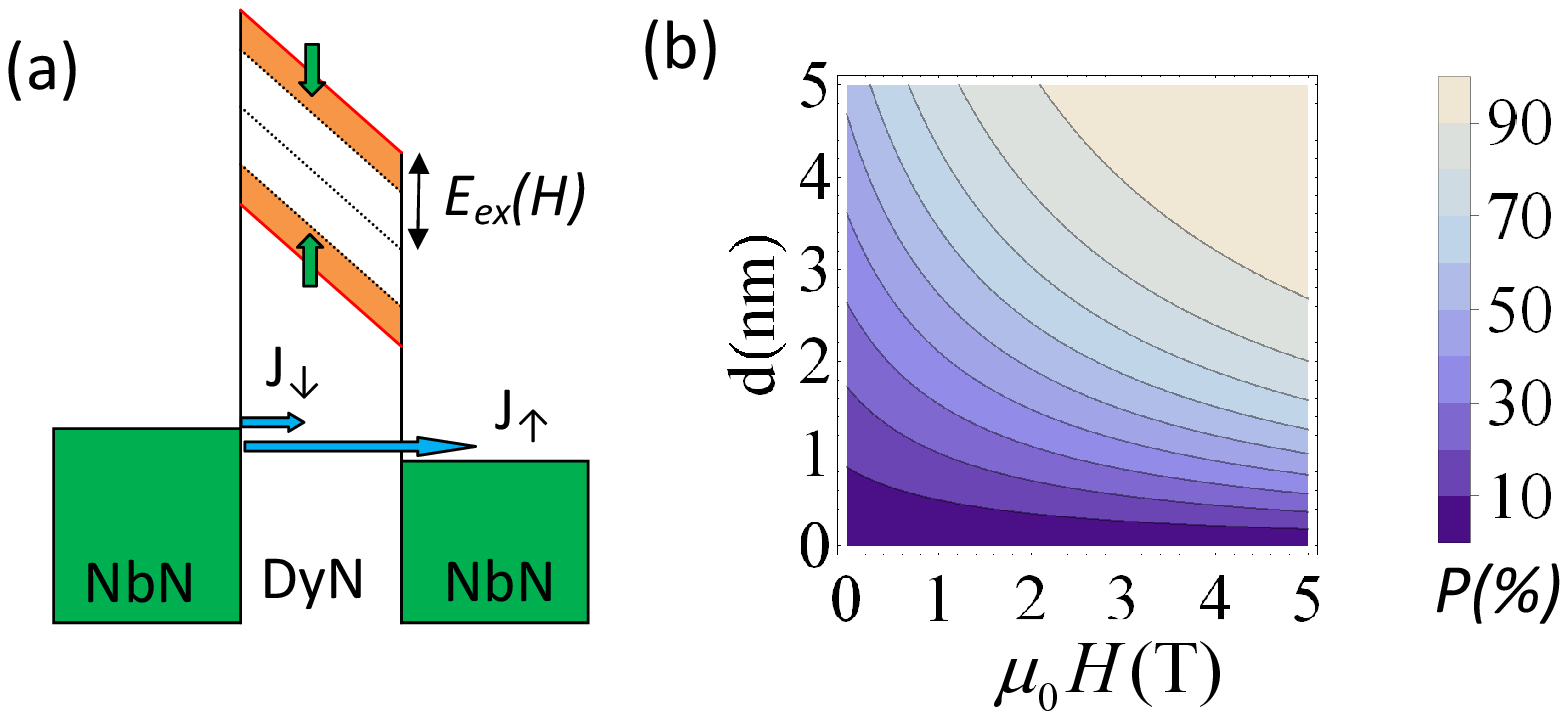}
\end{center}
\caption{\label{fig2}(Color online) (a) Schematic illustration of
the spin filter mechanism in presence of magnetic field. Spin-up
and spin-down bands are further split by magnetic field dependent
exchange splitting $E_{ex}(H)$ (shown in orange color). (b)
Expected spin-filter efficiency $P$ for different thickness $d$ of
the tunnel barrier with magnetic field $H$. The spin-filter
efficiency was calculated for a  barrier with height $\Phi _0$ =
60 meV and exchange splitting $E_{ex}(0)$ = 5.6 meV (see text for
details).}
\end{figure}

The tunneling nature of the junctions was confirmed through $I-V$
measurements at different temperatures.  Simmons
model\cite{simmons} was fitted to the $I-V$ curve  above
$T_{Curie}$ where exchange splitting of the barrier $E_{ex}$ = 0.
For DyN junction a barrier height $\Phi _0$ = 60 meV and width $d$
= 3.3 nm were found from the fitting at 50 K (see Supplementary
Information\cite{supp}). For the GdN junctions Simmons model
fitting was done at 100 K and  barrier height $\Phi _0$ = 111 meV
and width $d$ = 2.8 nm were found from the fitting. The $I-V$ and
$dI/dV-V$ measurement of both the DyN and GdN junctions at 4.2 K
showed clear superconducting gap structure. This confirms that the
electrical transport in our device is dominated by tunneling
process.

The mechanism of spin-filtering in presence of a magnetic field
can be understood with a tunneling model as shown in Fig. 2(a).
Below $T_{Curie}$, the conduction band is split by the
ferromagnetic exchange interaction leading to a lower barrier
height $\Phi _{\uparrow }$ = $\Phi _0 - E_{ex}$ for up-spin and
higher $\Phi _{\downarrow}$ = $\Phi _0 + E_{ex}$ for down-spin
electrons. The spin filtering efficiency is usually given by, $
P{\rm{ = }}\frac{{J_ \uparrow - J_ \downarrow }}{{J_ \uparrow + J_
\downarrow  }}$, where $J_ \uparrow$ and $J_ \downarrow$  are
spin-up and spin-down current density, respectively\cite{santos}.
As per the Wentzel-Kramers-Brillouin (WKB) approximation the
tunneling current density for each spin sign exponentially depends
on the relevant barrier height $\Phi _{\uparrow }$ (or $\Phi
_{\downarrow}$) and can be written as $J_{ \uparrow ( \downarrow
)}  \propto \exp \left( { - \frac{{2d}}{\hbar }\sqrt {2m \Phi _{
\uparrow ( \downarrow )} } } \right)$; where $d$ is the thickness
of the tunnel barrier and $m$ is the electron mass. In the
presence of a magnetic field the barrier height can further change
due to magnetic field dependence of exchange splitting $E_{ex}
(H)$. The additional spin splitting of the tunnel barrier due to
magnetic field is shown in Fig. 2(a). Considering magnetic field
dependency on barrier height, the spin filter efficiency at very
low applied bias voltage ($V$ $\rightarrow$0) can be written as;

\begin{equation}
P{\rm{ = }}\frac{{e^{{\rm{ - }}\frac{{{\rm{2d}}}}{\hbar }\sqrt
{2m[\Phi _0  - E_{ex} (H)]} }  - e^{{\rm{ -
}}\frac{{{\rm{2d}}}}{\hbar }\sqrt {2m[\Phi _0  + E_{ex} (H)]} }
}}{{e^{{\rm{ - }}\frac{{{\rm{2d}}}}{\hbar }\sqrt {2m[\Phi _0 -
E_{ex} (H)]} } {\rm{ + }}e^{{\rm{ - }}\frac{{{\rm{2d}}}}{\hbar
}\sqrt {2m[\Phi _0  + E_{ex} (H)]} } }}
\end{equation}

The exchange splitting $E_{ex}(H)$ can be calculated with a known
value of the spin-filter efficiency using Eq. [1]. At 11 K using
$P$ $\sim$37 $\%$ for the DyN tunnel junction we found
$E_{ex}(0)$ $\approx$5.6 meV for $H$ = 0 T.   Similar calculations
gave $E_{ex}(0)$ $\approx$47 meV at 11 K for the GdN tunnel
junctions. The spin-filter efficiency is usually determined by the
ratio $ E_{ex} /\Phi _0 $ which is only $\sim$0.09 for DyN tunnel
junction. In GdN tunnel junction $ E_{ex} /\Phi _0 $ $\sim$0.42.
For comparison, in EuO spin-filter junctions $ E_{ex} /\Phi _0 $
$\sim$0.3 which leads to a very high spin-filter efficiency
$\sim$98$\%$\cite{santos}. Figure 2(b) shows the calculated
spin-filter efficiency using Eq. [1] in the magnetic field range
$\mu_0H$ = 0 to 1 T for tunnel barrier in the thickness range 1 to
5 nm. The magnetic field dependence of exchange splitting was
assumed to be, $E_{ex} (H) = E_{ex} (0) + \alpha _s H $ for the
calculation of $P$. Here $\alpha _s$ is a temperature dependent
constant. Justification for such assumption is given in the
discussion section later on. For the calculation $\Phi _0 $ = 60
meV, $E_{ex}$ = 5.6 meV, and $\alpha _s $=0.004 was used. Clearly
very high spin-filter efficiency up to $\sim$100 $\%$ can be
obtained in thicker tunnel junctions with smaller magnetic field.
Therefore, magnetic field can be used to tune spin-filter
efficiency in our tunnel junctions.

\begin{figure}[!h]
\begin{center}
\abovecaptionskip -10cm
\includegraphics [width=9 cm]{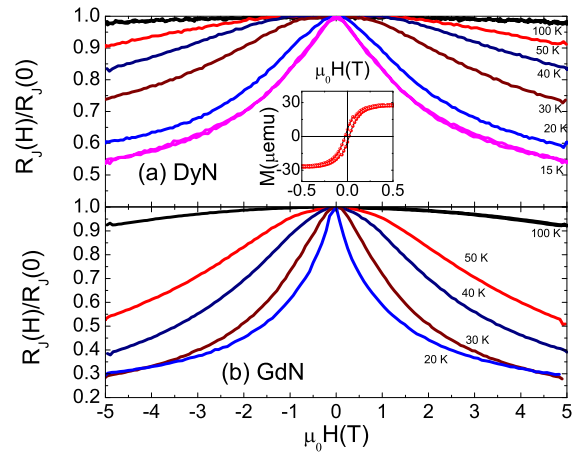}
\end{center}
\caption{\label{fig3}(Color online) Magnetic field dependence of
normalized junction resistance $R_N(H)$ (= $R_J(H)/R_J(0)$) of the
(a) NbN - DyN(3 nm)- NbN and (b) NbN-GdN(2.6 nm)-NbN tunnel
junction in the temperature range 15-100 K. Inset in Fig. 1(a)
shows magnetization $M(H)$ measurement at 13 K of a NbN-DyN(3
nm)-NbN trilayer film deposited at the same time.}
\end{figure}

Figure 3(a) and (b) show normalized junction resistance $R_N(H)$
(= $R_J(H)/R_J(0)$) at different temperatures for the DyN and GdN
tunnel junction, respectively. At 100 K negligibly small change in
$R_N(H)$ with magnetic field was found for both DyN and GdN tunnel
junction. As temperature is decreased $R_N(H)$ is found to
decrease rapidly with magnetic field. A much larger decrease
$\sim$$70 \%$ was found for the GdN junctions compared to $\sim$45
$\%$ in the DyN junction at the same temperature (20 K). The
$R_N(H)$ showed nonlinear dependence with magnetic field which
becomes prominent as temperature is decreased. As the junctions
were cooled down below $T_{Curie}$ the magnetization of the tunnel
barrier increases along with increase in spin-filter efficiency.
Therefore, the rapid decrease of $R_N(H)$ along with increase in
nonlinearity at lower temperature can be understood by considering
the magnetic field dependence of barrier heights $\Phi _{ \uparrow
( \downarrow )}$. Field dependence of magnetization of the
NbN-DyN(3 nm)-NbN trilayer film measured at 13 K is shown in the
inset of Fig. 3. The film was deposited at the same time as the
trilayer from which tunnel junction is fabricated.

\begin{figure}[!h]
\begin{center}
\abovecaptionskip -10cm
\includegraphics [width=9 cm]{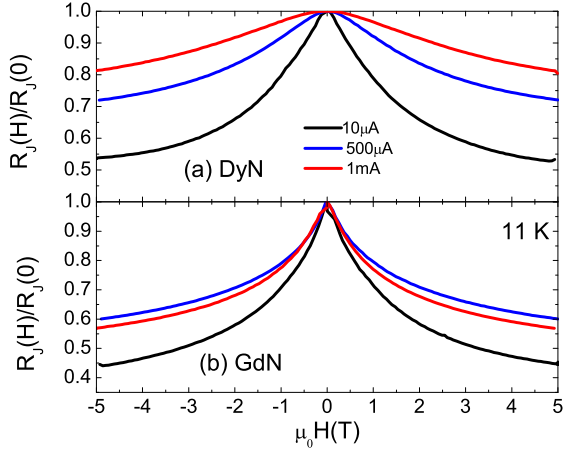}
\end{center}
\caption{\label{fig4} (Color online) Magnetic field dependence of
normalized junction resistance $R_N(H)$ (= $R_J(H)/R_J(0)$) of the
(a) NbN - DyN(3 nm)- NbN and (b) NbN-GdN(2.6 nm)-NbN tunnel
junction at 11 K with different bias current. Measurement was done
in a constant current mode with different current values.}
\end{figure}

We also measured the magnetic field dependence of the junction
resistance with different bias current. Fig. 4(a) and (b) shows
the field dependence of $R_N(H)$ measured at 11 K with bias
current $I$ = 10 $\mu$A to 1 mA. The measurements were done in a
constant current mode. We also found similar results with
measurements done in a constant voltage mode (see Supplementary
material). For the DyN junction at low bias current $\sim$10
$\mu$A we found $\sim$45 $\%$ decrease in $R_N(H)$ with a magnetic
field of 5 T. Larger $\sim$56 $\%$ decrease in $R_N(H)$ was found
for the GdN junction. As bias current was increased to $\sim$1 mA
the change in $R_N(H)$ was found to reduce significantly along
with suppression of nonlinearity. The DyN junctions showed much
stronger bias current dependence compared to GdN junctions.

\section{Discussion}
\begin{figure}[!h]
\begin{center}
\abovecaptionskip -10cm
\includegraphics [width=8 cm]{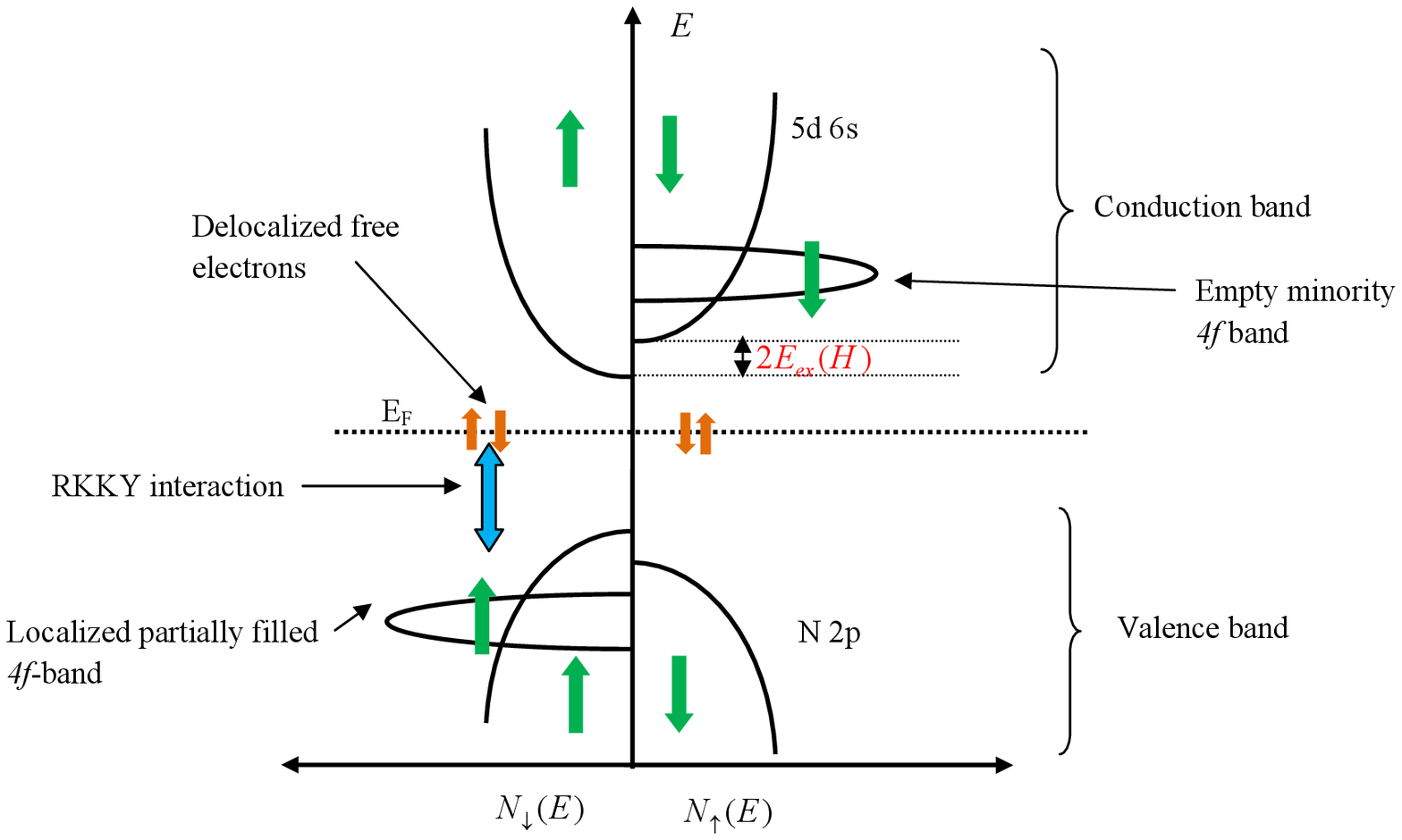}
\end{center}
\caption{\label{fig5}(Color online) Schematic of a simplified band
diagram of REN.\cite{trodhal,larson,mitra-band}}
\end{figure}

In spin-filter tunnel junctions difference in barrier height for
spin-up and down electrons appear because of the exchange
splitting of the conduction band of the ferromagnetic
semiconductor  below $T_{Curie}$. Therefore, a detail
understanding of the temperature and magnetic field dependence of
exchange splitting can be obtained from the band
structure\cite{Petukhov,Hasegawa}. Fig. 5 shows schematic of a
simplified band diagram of REN (more accurately for GdN and DyN).
In REN the valance band is usually formed from N 2p band which is
hybridized with RE 5d (6s) bands. While the conduction band
comprises RE 5d and 6s bands which is also hybridized with N 2p
bands. The majority RE $4f$ spin-up band is partially filled and
lies below the valance band maximum. The empty minority spin-down
$4f$ band lies above the conduction band minimum. As $4f$ bands
are highly localized with no direct orbital overlap, the $f-f$
band interaction between neighboring RE atoms is very week. The
magnetism in REN arises due to many competing direct and indirect
exchange interactions between $4f$ bands. Besides carrier mediated
Ruderman-Kittel-Kasuya-Yoshida (RKKY) type exchange interaction is
also proposed to exit in REN\cite{sharma}. The delocalized free
electrons close to Fermi level can mediate such interaction. The
magnetism in REN is quite complex compared to transition metal
ferromagnets where $d$ electrons are more itinerant in nature.
Magnetism in REN can be appropriately described by Heisenberg type
Hamiltonian
\begin{equation}
H =  - \sum\limits_{i,j} {J_{ij} S_i S_j },
\end{equation}
where $J_{ij}$ are exchange coupling constant between the
localized spin on the site $i$ and  $j$.  In a mean field
approximation the Curie temperature can be written
as\cite{pliu,smart}
\begin{equation}
k_B T_{Curie}  = (2/3)S(S + 1)\sum\limits_{m = 1}^N {z_m J_m },
\end{equation}
where $z_m$ is the number of nearest neighbor and $J_m$ is
corresponding exchange coupling constant. Here $S$ is total spin
on RE ion. The summation $N$ runs over all possible set of
neighbors. Considering exchange interaction between the nearest
and next-nearest neighbor only we can write, $\sum\limits_{m =
1}^N {z_m J_m }\approx (12J_1 + 6J_2 )$.  Here the nearest
neighbor exchange interaction $J_1$ involves virtual excitation of
majority $4f$ electron to the $5d$ band and subsequent exchange
interaction with neighboring RE $4f$ spins. The next nearest
neighbor superexchange interaction $J_2$ is  between two RE 5d
bands mediated by N 2p band. Here $J_1$ ($J_1
> 0$) is ferromagnetic (FM) and $J_2$ ($J_2
< 0$) is antiferromagnetic (AFM) coupling. The exchange coupling
constants $J_{1,2}$ have shown to strongly depend on lattice
constant and external magnetic field\cite{duan,Leuenberger}. The
AFM exchange $J_2$ can lead to a component of the total moment on
RE ions not aligned along net magnetization direction even below
$T_{Curie}$\cite{kartik-jpcm,sattler,demangeat}. Large external
magnetic field can strengthen FM exchange $J_1$ while weakening
AFM exchange $J_2$. According to Eq. (3) this will lead to clear
enhancement in $T_{Curie}$. Increase in $T_{Curie}$ as shown in
Fig. 1 strongly suggest suppression of AFM exchange with magnetic
field. Large magnetic field can ease randomly unaligned moments
line up along the net magnetization direction. This is similar to
paramagnetism. Therefore, phenomenologically the field dependence
of exchange splitting can be written as

\begin{equation}
E_{ex} (H) = E_{ex} (0) + E_{ex}^{sat} L(x),
\end{equation}
where $L(x) = \coth (x) - \frac{1}{x}$ is the Langevin function
and $x = \frac{{\mu _{eff}H}}{{k_B T}}$ is the ratio of magnetic
energy to thermal energy. Here $\mu _{eff}$ is the effective
moment on RE ion pointing away from net magnetization
direction\cite{note}. The exchange splitting  $E_{ex} (0)$ and
$E_{ex}^{sat}$ denote zero field and saturation ($H \to \infty$)
value, respectively. In the limit $x<<1$ (i.e, $\mu_{eff}H < < k_B
T$), Eq. (4) can be written as
\begin{equation}
E_{ex} (H) = E_{ex} (0 ) + \alpha_s H,
\end{equation}
where $\alpha_s = \frac{{\mu _{eff}}}{{3k_B T}}E_{ex}^{sat}$, is a
temperature dependent constant. Exponential dependence of
tunneling current on barrier height ensures that even a small
change in $E_{ex}(H)$ gets reflected in the tunneling conductance.

Magnetic property of Eu-chalcogenides and REN are analogues; in
both the cases ferromagnetism originates from competing indirect
exchange and superexchange interactions. In particular the
electronic band structure in EuO and GdN are surprisingly similar.
The only difference is the position of the $4f$ band in case of
EuO is close to Fermi level above the anion 2p band. Therefore,
the indirect exchange interactions in EuO is much stronger
compared to GdN. Paramagnetic surface sheets are usually observed
in EuO due to competing FM and AFM exchange
coupling\cite{sattler}. Magnetic field-enhanced spin-filtering has
also been observed in EuSe which is considered to be an
antiferromagnet at very low field and becomes ferromagnetic with
large magnetic field\cite{mooderaprl}.

We now try to explain the magnetic field and bias current
(voltage) dependence of the junction resistance using tunneling
model. Similar to Mott's two-current model\cite{mott}, we assume
the spin currents for spin-up and spin-down electrons are
independent of each other. Therefore, the total conductance of our
tunnel junction can be written as, $G(V) = G_ \uparrow (V)   + G_
\downarrow (V)$, where $G_ \uparrow (V)$ and $G_ \downarrow (V)$
are conductances for up-spin and down-spin electrons,
respectively. In the low voltage limit ($V \rightarrow0$) the
conductances $G_ \uparrow (0)$ and $G_ \downarrow (0)$ can be
calculated from Simmons' model\cite{simmons} as:

\begin{equation}
 G_{ \uparrow ( \downarrow )}(H)  = \frac{{3\sqrt {2m\Phi _{
\uparrow ( \downarrow )} } }}{{2d}}\left( {\frac{e}{h}} \right)^2
\exp ( - \frac{{2d}}{\hbar }\sqrt {2m\Phi _{ \uparrow ( \downarrow
)} } ),
\end{equation}

where $m$ is mass of the electron, $\hbar$ is the Planck's
constant and $d$ is the barrier width. For simplicity of
calculation following Eq. (5), the barrier height for spin-up and
spin-down electrons in presence of magnetic field can be written
as; $ \Phi _ \uparrow = \Phi _0 - E_{ex}(0) - \alpha _s H$ and $
\Phi _ \downarrow = \Phi _0 + E_{ex}(0) + \alpha _s H$,
respectively. Calculated normalized junction resistance $R_N(H)$ =
$\frac{{G(0)}}{{G(H)}}$, for the DyN tunnel junction using Eq. (6)
with $\Phi _0 $ = 60 meV and $E_{ex} (0)$ = 5.6 meV is shown in
Fig. 6. Clearly one can see that as $\alpha _s$ increases $R_N(H)$
decrease rapidly and becomes nonlinear with magnetic field. The
temperature dependence of $R_N(H)$ shown in Fig. 3 is consistent
with increase in $\alpha _s$ with decreasing temperature.
Although, our model provides a qualitative picture, for more
accurate estimation of $R_N(H)$ one has to consider bias voltage
dependence of $G(V)$ and exact functional dependence of
$E_{ex}(H)$ with magnetic field $H$. In our model, we have
considered only direct tunneling process, however, other charge
transport mechanism like sequential tunneling and inelastic
cotunneling through defects in the barrier can also contribute to
$R_N(H)$. Spin-flip scattering processes due to such charge
transport mechanism can lead to significant reduction in $R_N(H)$
calculated from our ideal model. Smaller $R_N(H)$ for higher bias
current as shown in Fig. 4 suggests possibility of a
magnetoelectric coupling within the barrier. The magnetoelectric
coupling can arise in our tunnel junctions due to double Schottky
nature of these barriers. Earlier we have observed an electric
field dependent spin-filter efficiency in NbN-GdN-NbN tunnel
junctions\cite{pal}.

\begin{figure}[!h]
\begin{center}
\abovecaptionskip -10cm
\includegraphics [width=8 cm]{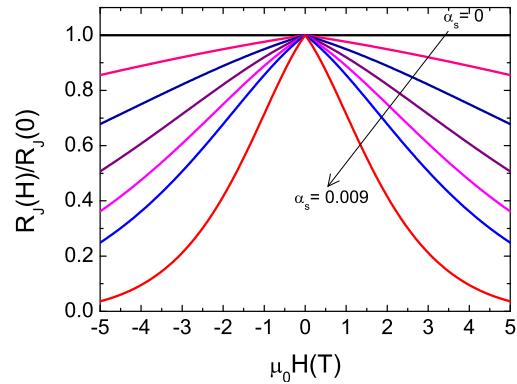}
\end{center}
\caption{\label{fig5} (Color online) Calculated normalized
junction resistance $R_N(H)$ as a function of magnetic field for
$\alpha_s$ = 0 to 0.009 (see text for details).}
\end{figure}

\section{Conclusions}
In conclusion, we have fabricated spin-filter tunnel junctions
with DyN and GdN tunnel barrier sandwiched between two
superconducting NbN electrodes. We have made an extensive study on
the effect of applied bias current and magnetic field on the
spin-filtering efficiency. Using the Simmons model we have
calculated barrier height $\Phi _0$ $\approx$60  and 111 meV for
tunnel junction with DyN and GdN barriers, respectively. In DyN
tunnel junctions,  a small exchange splitting $ E_{ex}(0)$
$\approx$5.6 meV of the barrier height at 11 K resulted in
$\sim$37$\%$ spin-filter efficiency. Larger $ E_{ex}(0)$
$\approx$46 meV lead to a huge $\sim$97$\%$ spin-filter efficiency
in the case of GdN junctions. In DyN junctions the spin filter
efficiency is found to increase to $\sim$87$\%$ when magnetic
field $H$ = 5 T was applied. Similar field-induced increase in
spin-filter efficiency was also found for GdN junctions. The
increase in spin-filter efficiency can be understood by
considering further lowering of the up-spin barrier height by
magnetic field-induced increase in $E_{ex}(H)$. Magnetic
field-induced enhanced ferromagnetism in these junctions were
supported by observation of rise in $T_{Curie}$ with magnetic
field in $R(T)$ measurements. Our findings are potentially useful
for the control of spin polarized super-current in spintronics
devices

\noindent  \textbf{Acknowledgments}\\

This work was supported by the ERC Advanced Investigator Grant
SUPERSPIN. XLW acknowledges funding from the NSFC (Grant No.
11404323).

\newpage
\begin{widetext}
\textbf{Supplementary Information}

I-V measurements of our tunnel junctions were done at different
temperatures. Tunnel barrier height was obtained by fitting IV
curve to the Simmons model. For a rectangular barrier in the
voltage range $0 < V < \Phi _0$, Simmons model expressed in
practical units can be written as;

\begin{equation}
J(V) = \frac{{6.2 \times 10^{10} }}{{d^2 }} \biggl[ {\biggl( {\Phi
_0 - \frac{V}{2}} \biggr)\exp ( - 1.025d\biggl( {\Phi _0  -
\frac{V}{2}} \biggr)^{1/2} - \biggl( {\Phi _0  + \frac{V}{2}}
\biggr)\exp ( - 1.025d\biggl( {\Phi _0  + \frac{V}{2}}
\biggr)^{1/2} } \biggr]
\end{equation}
Here $J(V)$  is expressed in A/cm$^2$, $ \Phi _0$ in V and $d$  in
{\AA} unit. Fig. 7 and 8 shows Simmons model fitting to IV curve
measured above the Curie temperature $T_{Curie}$ for the NbN -
DyN(3 nm)- NbN and NbN-GdN(2.6 nm)-NbN tunnel junction,
respectively.
\begin{figure}[b]
\begin{center}
\abovecaptionskip -10cm
\includegraphics [width=8 cm]{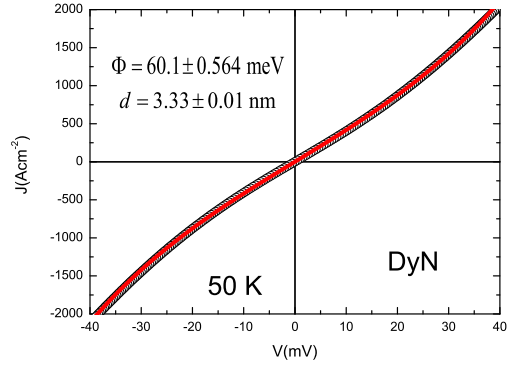}
\end{center}
\caption{\label{fig5}(Color online) IV-curve measured at 50 K for
the NbN - DyN(3 nm)- NbN tunnel junction. Solid red line shows
fitting to the  Simmons model.}
\end{figure}

\begin{figure}[H]
\begin{center}
\abovecaptionskip -10cm
\includegraphics [width=8 cm]{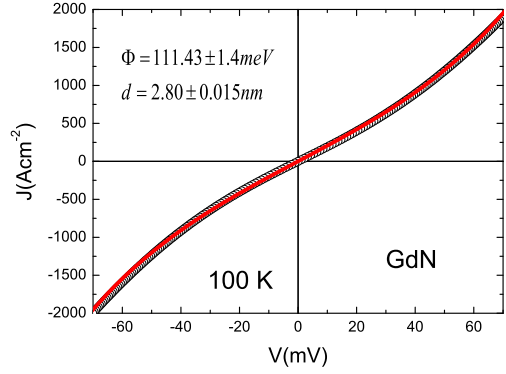}
\end{center}
\caption{\label{fig5}(Color online) IV-curve measured at 100 K for
the NbN-GdN(2.6 nm)-NbN tunnel junction. Solid red line shows
fitting to the  Simmons model.}
\end{figure}
We also measured I-V and $dI/dV-V$ of the tunnel junctions below
the critical temperature $T_C$ of NbN.  Fig. 9 and Fig. 10 shows
normalized conductance spectra and I-V of the same junctions
measured at 4.2 K. Superconducting gap structure with $2\Delta
\approx$2.88 meV can be seen.
\begin{figure}[H]
\begin{center}
\abovecaptionskip -10cm
\includegraphics [width=8 cm]{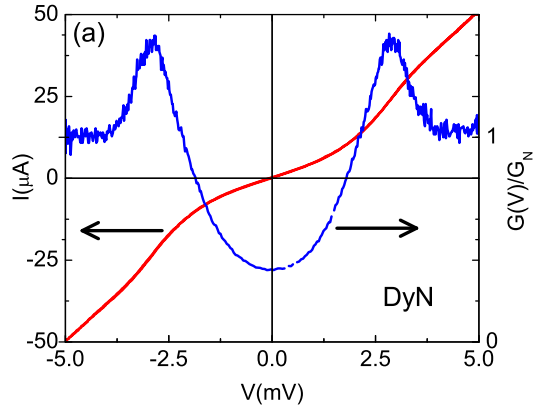}
\end{center}
\caption{\label{fig5}(Color online) Normalized conductivity
$G(V)-V$ and I-V curve measured at 4.2 K for the NbN - DyN(3 nm)-
NbN tunnel junction}
\end{figure}

\begin{figure}[!h]
\begin{center}
\abovecaptionskip -10cm
\includegraphics [width=8 cm]{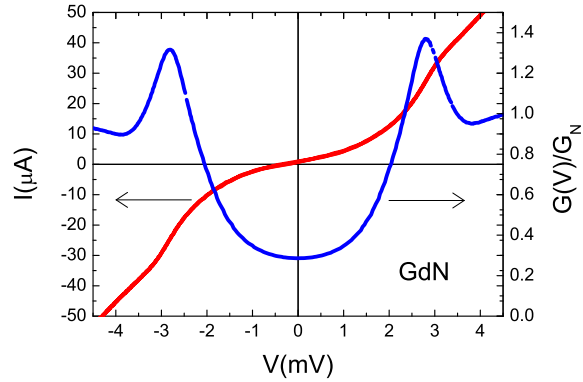}
\end{center}
\caption{\label{fig5}(Color online) Normalized conductivity
$G(V)-V$ and I-V curve measured at 4.2 K for the NbN-GdN(2.6
nm)-NbN tunnel junction}
\end{figure}
The I-V measurement shown in Fig.8 and Fig. 9 clearly imply that
the electrical transport in our device is tunneling type.

\begin{figure}[H]
\begin{center}
\abovecaptionskip -10cm
\includegraphics [width=8 cm]{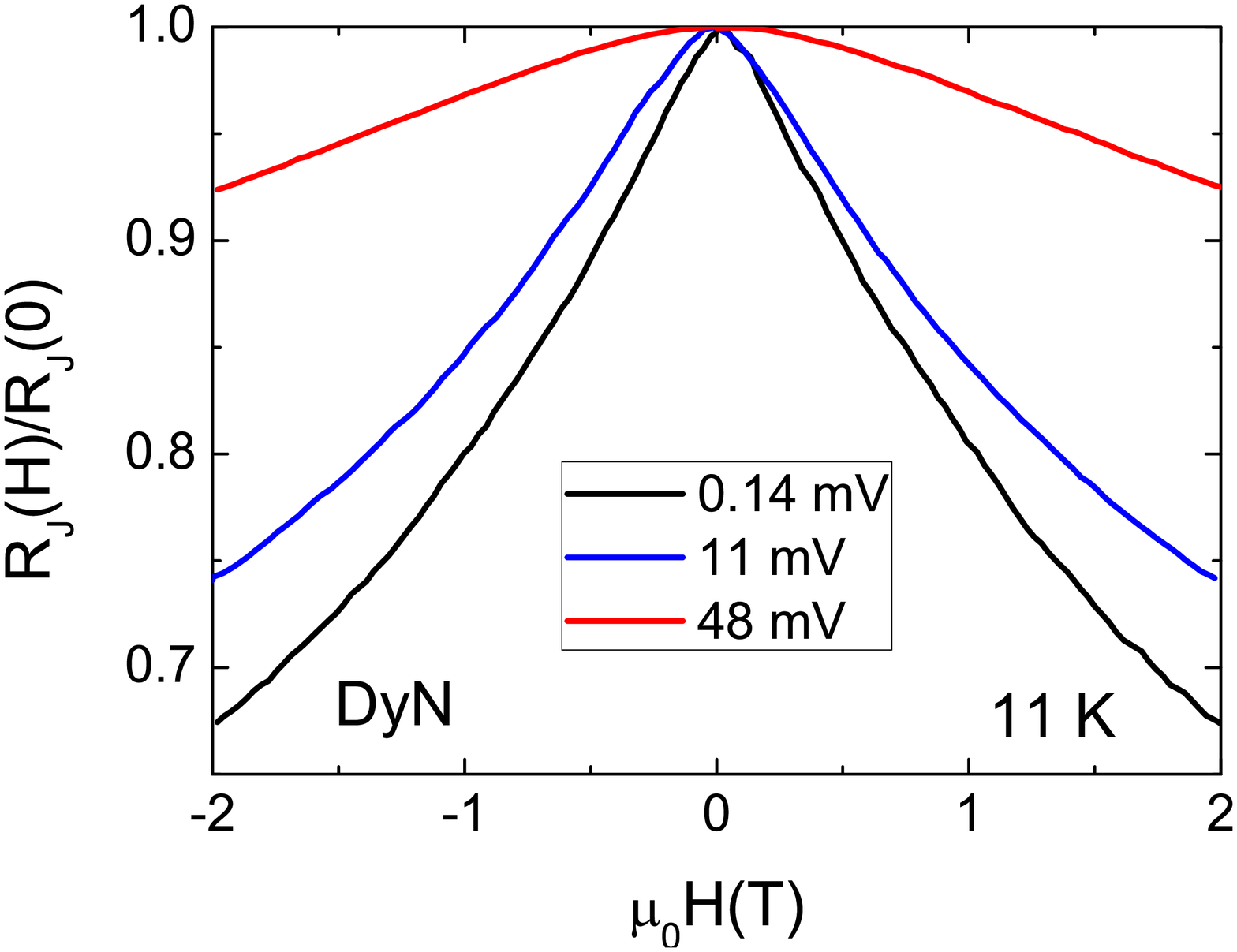}
\end{center}
\caption{\label{fig5}(Color online) Magnetic field dependence of
normalized junction resistance   of the NbN - DyN(3 nm)- NbN
tunnel junction measured with different bias voltage in constant
voltage mode.}
\end{figure}
\end{widetext}


\begin{thebibliography}{99}
\bibitem{note}
The effective moment  $\mu _{eff}$   is much smaller than the
total moment on RE ion; $ \mu = g\mu _B S $.  Only a small
component of total moment on RE ion can point against net
magnetization direction below T$_{Curie}$ due to AFM exchange
interaction. Here $g$ is Land\'{e} g-factor and $S$ is total spin
on RE ion.
\end{thebibliography}
\end{document}